# Spintronics of Organometal Trihalide Perovskites

Dali Sun†, Chuang Zhang†, Marzieh Kavand†, Kipp J. van Schooten, Hans Malissa, Matthew Groesbeck, Ryan McLaughlin, Christoph Boehme*, and Z. Valy Vardeny*

*Department of Physics & Astronomy, University of Utah, Salt Lake City, Utah, 84112, USA*

**The family of organometal trihalide perovskite (OTP), $CH_3NH_3PbX_3$ (where X is halogen) has recently revolutionized the photovoltaics field and shows promise in a variety of optoelectronic applications. The characteristic spin properties of charge and neutral excitations in OTPs are influenced by the large spin-orbit coupling of the Pb atoms, which may lead to spin-based device applications. Here we report the first studies of pure spin-current and spin-aligned carrier injection in OTP spintronics devices using spin-pumping and spin-injection, respectively. We measure a relatively large inverse-spin-Hall effect using pulsed microwave excitation in OTP devices at resonance with a ferromagnetic substrate, from which we derive room temperature spin diffusion length, $\lambda_{sd} \sim 9$nm; and low-temperature giant magnetoresistance in OTP-based spin-valves from which we estimate $\lambda_{sd} \sim 85$nm.**

†These authors contributed equally to this work.
*To whom correspondence should be addressed: boehme@physics.utah.edu, val@physics.utah.edu

The hybrid organometal trihalide perovskites (OTP) such as MAPbX$_3$, where MA is CH$_3$NH$_3$ and X is halogen, have emerged as a new class of solution-based photovoltaic materials that have revolutionized the photovoltaic solar cells research field, having power conversion efficiencies reaching ~20%[1-3]. Owing to the low-temperature processing[4,5], the simplicity of thin-film device fabrication[6-9], and electronic property tunability that originates from various synthesis routes[2,3,10], OTP materials have also been considered in other optoelectronic applications. For example, these compounds exhibit excellent optical gain for lasing[11] and light-emitting[12-14] properties. Recently, it has been predicted that the OTP semiconductors possess strong spin-orbit coupling (SOC)[15], and Rashba-type splitting[16] that originate from the heavy metal (Pb); this implies that OTP may also be considered for spintronics applications.

Recent studies of spin-related phenomena in hybrid perovskites have been focused on the magnetic field modulation of the photocurrent and electroluminescence in OTP-based optoelectronic devices, and photoluminescence in OTP films[17,18]; as well as photo-induced magnetism using the circularly polarized picosecond pump-probe technique[19]. However, OTP spintronics devices that show spin-current carried by spin-polarized carriers 20 are still missing[20]. Here we report on spin-current in MAPbI$_3$ and MAPbI$_{3-x}$Cl$_x$ films induced by both spin-pumping and spin-injection from ferromagnetic (FM) substrates. For the spin-pumping experiments, the spin-current is detected by the inverse spin Hall effect (ISHE) that occurs when the FM substrate is in resonance with pulsed microwave (MW) excitation[21,22]. From the ISHE value, we estimate the SOC strength and spin diffusion length in these materials[23]. The spin aligned carrier injection is demonstrated in OTP-based spin-valve (SV) devices that exhibit substantial magnetoresistance (MR) response up to ~6% at low temperatures.

Firstly, we demonstrate spin-pumping from FM substrate into OTP films using four different devices. Fig. 1a (upper panel) shows a schematic illustration of the pulsed ISHE (p-ISHE) device used in this study[23]. Spin-pumping occurs when the FM metal (here Ni$_{80}$Fe$_{20}$, NiFe), used as a substrate on which the OTP film is deposited, exhibits strong FM-resonance (FMR) that induces a spin-current into the nonmagnetic OTP layer (see Fig. 1a, lower panel)[24-28]. To increase the sensitivity of our apparatus we used a *pulsed* (2 μs duration) MW that induces transient precession of the FM magnetization vector, **M**; which, in turn, generates magnons in the FM of which scattering from the FM/OTP interface launches a pulsed spin-current ($J_S$) into the hybrid perovskite film. Due to the existence of strong SOC in the OTP film, the spin-current is converted into charge-current, $J_C$ via the ISHE described by[25, 28]:

$$J_C \sim \sigma E_{ISHE} \propto \theta_{SHE}\left(\frac{2e}{\hbar}\right) J_S \times S \tag{1}$$

where $E_{ISHE}$ is the SOC-induced electric field between the two Cu contacts, having direction perpendicular to the spin-current; and $\theta_{SHE}$, e, $\hbar$, $S$, and σ denote, respectively the spin Hall angle, elementary charge, Planck constant, spin polarization vector and the OTP conductivity along the FM/OTP interface. Using the p-ISHE technique allows for the application of much stronger MW peak power (~1kW) that in turn generates a much stronger ISHE signal compared to continuous wave (cw) MW excitation with typical powers of ~200 mW; while, at the same time, it greatly suppresses possible device heating effects due to low duty cycle[29].

A clear manifestation of spin-pumping into OTP films may be obtained from the increase, Δα, of the Gilbert damping constant, α, when FMR is measured on NiFe/OTP bilayers[30]. Fig. 1b shows the cw FMR($B$) response measured on NiFe/OTP bilayers at various MW frequencies, $f$, using a planar transmission line (see S.I. section A). The clearly observed increase in the FMR linewidth (measured by peak-to-peak difference field, $\Delta H_{pp}$) compared to that of NiFe film alone (see Fig. 1c inset), is a consequence of spin-pumping from the FM into the OTP film[28]. We note that $\Delta H_{pp}$ increases linearly with $f$ ($\propto \frac{4\pi\alpha f}{\sqrt{3}\gamma}$, where γ is the gyromagnetic ratio)[30], and consequently both α and Δα can be more precisely determined (Fig. 1b inset). It is clear from these experiments that there is excess damping in the NiFe/OTP junction related to the loss caused by spin-pumping into the OTP overlayer.

Fig. 1c shows the p-ISHE($B$) response of a NiFe/MAPbI$_3$ (10 nm)/Cu device at 1kW pulsed MW excitation power ($P_{MW}$), which exhibits a polarity change when measured at two opposite $B$ directions. p-ISHE($B$) response that follows the FMR($B$) response (Fig. 1c inset) is observed around the resonant field ($B_{res}$). The ISHE current observed here may be due to a combination of the ISHE in the OTP film, anomalous Hall effect (AHE) in the NiFe film, and heating effects (see discussion in S.I. section C)[25, 28]. The AHE($B$) contribution can be readily separated from that of ISHE($B$) by fitting the measured data with a combination of symmetric (pure ISHE($B$) response) and asymmetric (dispersion shape, AHE($B$)) functions[28]. To rule out heating effects, we note that its magnetic field response is independent of the **B** direction, in sharp contrast with the data (S.I. Fig. S3)[28].

The second device that demonstrates spin-pumping into OTP films is NiFe/MAPbI$_{3-x}$Cl$_x$ (67 nm)/Cu structure of which p-ISHE($B$) response is shown in Fig. 1d. Remarkably, the p-ISHE polarity in MAPbI$_{3-x}$Cl$_x$ is opposite to that of MAPbI$_3$. Chlorine incorporation has been shown to affect the OTP film morphology, which results in improved optoelectronic characteristics properties. In addition, the chlorine atoms mainly decorate the surfaces of the hybrid perovskite film[33]; this is consistent with our observation, since the actual p-ISHE response here occurs very close to the FM/OTP interface (S.I. Fig. S4).

P-ISHE responses were also measured in trilayer devices (NiFe/MAPbI$_3$/Pt and NiFe/MAPbI$_{3-x}$Cl$_x$/Pt), as shown in Fig. 1e and 1f, respectively. In these cases, the induced OTP spin-current diffuses (and attenuates) through the hybrid perovskite film, reaching the thin Pt film deposited underneath, where it converts into charge-current via the ISHE of the Pt layer[27]. Because of the large $\theta_{SHE}$ (~$10^{-2}$) in Pt, p-ISHE is now one to two orders of magnitude larger than that obtained in the pure OTP film. Consequently the trilayer device can be used as a sensitive detector of spin transport in the perovskite film, thereby allowing to extract the spin diffusion length, $\lambda_{sd}$ of the OTP[27] film, as shown in Fig. 2b (see below and S.I. Fig. S5).

The successful demonstration of ISHE response in OTP spin-related devices motivated the study of the spin-pumping characteristics at various MW intensities, OTP thicknesses, and temperatures, as shown in Fig. 2. The MW power dependence of the obtained p-ISHE response for a NiFe/MAPbI$_3$/Cu device at room temperature is shown in Fig. 2a, and summarized in Fig. 2a inset. It is clear that the p-ISHE current increases linearly with $P_{MW}$, as expected for an unsaturated ISHE response[28]. Fig. 2b shows the p-ISHE($B$) response measured in trilayer NiFe/MAPbI$_{3-x}$Cl$_x$/Pt devices at various perovskite thickness values, $d_N$. In the Pt film, $V_{pISHE}$ shows a monotonic decrease with $d_N$, which is well fitted by an exponential decay, $V_{pISHE}(d_N) \propto \exp(-d_N/\lambda_{sd})$[27].

From the exponential $V_{pISHE}(d_N)$ fit, we obtain $\lambda_{sd}$ ~9nm in MAPbI$_{3-x}$Cl$_x$ at 300K. Using the same procedure, we also obtain $\lambda_{sd}$ ~8nm in MAPbI$_3$ (S.I. Fig. S5).

The spin transport studies allow us to estimate the spin Hall angle ($\theta_{SHE}$) in the OTP layer. The spin-pumping related ISHE current, $I_C$ in the OTP layer can be expressed as[28]:

$$I_C = l\theta_{SHE}\left(\frac{2e}{\hbar}\right)\lambda_{sd}\tanh\left(\frac{d_N}{2\lambda_{sd}}\right)j_S^0, \qquad (2)$$

where $l$ is the geometrical device width and $j_S^0$ is the spin-current in the OTP layer perpendicular to the FM/NFM interface. $j_S^0$ may be obtained from the extra attenuation of the FMR response that leads to $\Delta\alpha$, as discussed above (see S.I. section E). We measured $I_C$ by taking into account the device structure, detection electronics, and the AC response of each electronic component during the pulsed measurement[23]. From the measured $I_C$, $\lambda_{sd}$ obtained above, and FMR($B$) response, we calculated the spin Hall angles for MAPbI$_{3-x}$Cl$_x$ ($\theta_{SHE} \approx +6.0\times 10^{-4}$) and MAPbI$_3$ ($\theta_{SHE} \approx -1.5\times 10^{-3}$). Both values are larger than $\theta_{SHE}$ in doped GaAs, Si, and the conducting polymer PEDOT:PSS[24-26]. This relatively large $\theta_{SHE}$ value is in agreement with the strong SOC of the Pb atoms in OTP, as predicted by theory[15,16].

The temperature dependence of p-ISHE($B$) response in NiFe/MAPbI$_3$(10nm)/Cu is presented in Fig. 2c and summarized in Fig. 2d, where it is normalized by taking into account the FMR amplitude, FMR linewidth, and device resistance. Based on Eq. (S1), the normalized p-ISHE response is proportional to the product $\lambda_{sd}\theta_{SHE}$[23], which is strongly dependent on the temperature (Fig. 3d). $\lambda_{sd}\theta_{SHE}$ dramatically increases for $T$<150K, whereas for T>150K it substantially decreases with $T$; at $T$~150K the product $\lambda_{sd}\theta_{SHE}$ shows a marked discontinuity due to the known structural phase transition of MAPbI$_3$ from tetragonal to orthorhombic crystal structure[34]. We also measured the temperature dependence of $\lambda_{sd}\theta_{SHE}$ in three other OTP devices, namely NiFe/MAPbI$_3$(5nm)/Cu, NiFe/MAPbI$_3$(5nm)/Pt, and NiFe/MAPbI$_{3-x}$Cl$_x$ (32nm)/Pt (S.I. Fig. S6). In all of these devices, $\lambda_{sd}\theta_{SHE}$ shows the abrupt discontinuity at $T$~150K. Assuming that $\theta_{SHE}$ is temperature independent (since the SOC is weakly temperature dependent), the observed phase transition revealed by the $\lambda_{sd}\theta_{SHE}(T)$ response indicates that the spin diffusion length is different in the two structural phases. Since $\lambda_{sd} = \sqrt{3D\tau_s}$, where $\tau_s$ is the spin relaxation time and $D$ is the carrier diffusion coefficient ($\propto k_B T \mu / e$, with $\mu$ being the mobility), we thus conclude that $\mu$ shows a discontinuity at $T\sim150K$. This is reasonable since the phase transition is likely inhomogeneous, leading to a substantial increase of native defects, such as vacancies and dislocations for $T<150K$, which, in turn reduces the carrier mobility.

We have also succeeded in launching spin-current in OTP films by injecting spin-polarized carriers from FM electrodes, while detecting it using a second FM electrode in a device that is known as a spin-valve (SV). Our fabricated perovskite SV (OTP-SV) device (Fig. 3a) consists of an OTP interlayer (MAPbI$_{3-x}$Cl$_x$) sandwiched between a 100nm-thick La$_{0.67}$Sr$_{0.33}$MnO$_3$ (LSMO) FM electrode and a 10nm-thick cobalt electrode, each having a different coercive field, $B_{c1}=4$ mT and $B_{c2}=150$ mT, respectively[35-37] (see Fig. 3c). The spin-polarized carriers are electrically injected from the LSMO electrode into the OTP layer, and detected by the Co electrode using the change in the device resistance with $B$, which leads to magneto-resistance, $MR$ (see Fig. 3a). Upon sweeping the field $B$, the relative magnetization direction of the two FM electrodes change from the parallel (↑↑) (*P*) to the antiparallel (↑↓) (*AP*) configuration, and vice-versa, since $B_{c1} \neq B_{c2}$. Consequently, the device resistance, $R$ changes from $R_P$ to $R_{AP}$, causing the *MR(B)* response; it can be characterized by the figure-of-merit, $MR_{max}$ using the relation[37]:

$$MR_{max} = \left(\frac{R_{AP}-R_P}{R_P}\right)_{max} = \frac{2P_{LSMO}P_{Co}e^{-d_N/\lambda_{sd}}}{1+P_{Co}e^{-d_N/\lambda_{sd}}} \approx 2P_{LSMO}P_{Co}e^{-d_N/\lambda_{sd}}, \qquad (3)$$

where $P_{LSMO}$ (~95% at low temperature[36]) and $P_{Co}$ (~35%, depending on the environment) are the nominal spin injection polarization degrees of the LSMO and the Co electrode, respectively.

Fig. 3b shows a typical nonlinear *I-V* characteristic response of OTP-SV device at 10K and 300K, respectively; as well as a d*I*/d*V-V* response that is free from 'zero-bias anomaly'[38,39], which implies that charge carriers are electrically injected into the OTP layer without short-circuits such as pinholes. Since the OTP film is known to be a semiconductor with excellent photovoltaic properties, we were able to validate the pinhole-free charge transport through the OTP interlayer using the photocurrent generated upon device illumination (Fig. 3b). It is seen that the photocurrent increases by about a factor of two with respect to the dark current, even though the two electrodes work functions are almost equal (implying a small internal electric field in the device) (see S.I. Fig. S7). This effect is a direct proof that drift-current rather than tunneling occurs in the hybrid perovskite layer of the OPT-SV device.

It is possible to obtain the in-situ magnetization field response, M($B$) of the two FM electrodes while the OTP-SV device is biased. For this endeavor, we measured the electrode magneto-optical Kerr effect (MOKE) response using a home-built Sagnac interferometer with 20 nrad polarization rotation resolution[40]. The OTP-SV devices used for these experiments had a semitransparent Co electrode, which allowed the Sagnac beam to readily reach the LSMO electrode beneath the OTP layer. Fig. 3c (inset) shows the polarization rotation angle, $\theta(B)$ of the FM electrodes at 10K, where the Sagnac beam was focused outside and inside the device area, respectively. Although the thin Co electrode shows an unsaturated $\theta(B)$ response due to its unusual small thickness on top of OTP film, a clear two-step response of $\theta(B)$ that consists of a 'soft' component (LSMO) and 'hard' component (Co) is observed through the OTP-SV device. Therefore $\theta(B)$, which is related to the surface magnetization, $M_s(B)$ could be obtained at different temperatures, $T$ (Fig. 3d). Strikingly, distinct from $M(T)$ in the bulk LSMO electrode, which usually persists up to ~300K[37], $M_s$(LSMO) in the device configuration steeply decreases with $T$ and disappears at $T$>120K; whereas $B_c$(LSMO) is almost $T$-independent (Fig. 3d inset). The strong $M_s(T)$ decrease for the LSMO electrode has a dramatic effect on the OTP-SV device operation at high temperatures[37], as seen below.

Fig. 4a shows typical $MR(B)$ response of LSMO/MAPbI$_{3-x}$Cl$_x$(245nm)/Co/Al OTP-SV device measured at various bias voltages, $V$. We obtain a pronounced SV-related $MR(B)$ response having $MR_{max}$ within the two switching fields B$_1$≈4 mT and B$_2$≈100 mT, which is consistent with the coercive fields obtained by the Sagnac-measured MOKE through the device (Fig. 3c). We note that R$_{AP}$ > R$_P$, and consequently the measured $MR > 0$ is in agreement with conventional inorganic SV devices, but opposite than that in organic SV (OSV) devices[35-37]. The obtained $MR_{max}(V)$ response steeply decreases with $V$ (Fig. 4b), in agreement with OSV devices; this may indicate that it originates from the same mechanism as in OSV that gives a power law $MR_{max}$ decay ($MR_{max}$~$V^{-p}$)[37]. Importantly, we used the Sagnac interferometer to measure the MOKE response vs. *V*, and found that it is robust. This indicates that the spin-polarization degree, *P* of the FM electrodes does not diminish with *V*, in contrast with the steep $P_{LSMO}$ decrease with *T*.

Using Eq. (3) with $MR_{max}$ = 4%, $d_N$=245 nm, and $P_{LSMO}P_{Co}\approx$35%, we calculate a spin diffusion length $\lambda_{sd}\approx$85nm for MAPbI$_{3-x}$Cl$_x$ at 10K. This is in fair agreement with $\lambda_{sd}$ at 4K measured by ISHE (Fig. 2d) given the uncertainty in $P_{Co}$, which is known to change substantially when deposited on different materials. From the obtained $\lambda_{sd}$, we can now estimate the spin relaxation time, $\tau_S$, in MAPbI$_{3-x}$Cl$_x$ at low *T*. Using a carrier mobility of 100 cm$^2$/Vs, the Einstein relation between carrier mobility and the diffusion constant, and the relation $\lambda_{sd}=(3D\tau_S)^{1/2}$ that describes the spin diffusion process carried by the charge carriers, we calculate $\tau_S\approx$1 ns. This is in very good agreement with ps transient photoinduced Kerr effect measurements using the polarized pump-probe technique (see S.I. section I).

Fig. 4c presents a series of *MR(B)* responses measured at different temperatures of the same OTP-SV device; the extracted $MR_{max}$ vs. *T* is summarized in Fig. 4d. As is clearly seen, $MR_{max}(T)$ sharply decreases at high temperatures. This may originate from several mechanisms: (1) decreased $P_{Co}$ (weak *T* dependence); (2) decreased $M_s$ at the interface between LSMO and perovskite (strong *T* dependence, Fig. 3d) that leads to $P_{LSMO}$ decrease[37]; (3) decreased spin diffusion length due to the dominant SOC in the OTP layer. Indeed, when using Eq.(3), taking into account the three different *T*-dependences, namely $P_{Co}(T)$ and $P_{LSMO}(T)$ (from Fig. 3d), and $\lambda_{sd}(T)$ (from Fig. 2d), we obtain a good fit to the measured $MR_{max}(T)$, as seen in Fig. 4d.

In summary, we report spin-pumping and spin-injection studies in organometal trihalide perovskites. From the magnitude of the pulsed ISHE measurements, we estimate the SOC strength and spin diffusion length in the OTP film. We find that the spin Hall angle $\theta_{SHE}\approx$10$^{-3}$ in MAPbI$_3$ is only about one order of magnitude weaker than that in Pt (which is the standard ISHE metal used for spintronics applications); in agreement with the short spin diffusion length (~9 nm) in this compound at RT. Our MR measurements in OTP-SV devices hold promise for spin-related light emitting diodes based on the OTP materials, thus paving the way towards hybrid perovskite spintronics applications.

## Methods:

### (i) OTP-ISHE device fabrication

For the OTP-ISHE devices[23], two Al electrodes (150 nm) were grown on glass templates (3×50 mm) by sputtering and a wet-etch using conventional optical lithography. Subsequently, two Cu contacts with a gap of 50 μm (extended from an Al bottom electrode) were grown by e-beam evaporation through a shadow mask in a glove box integrated vacuum deposition chamber (Angstrom Engineering Inc.), having a base pressure of 3×10$^{-8}$ Torr. For the solution-based OTP-ISHE devices (see S.I. section A), the fabricated templates were transferred into a second glove box through an antechamber that is devoted to spin coating of the OTP layer. For the co-deposited OTP layer (see S.I. section A), the templates were transferred into the second organic deposition chamber for thermal deposition without breaking the vacuum. After OTP layer preparation as described above, Ni$_{80}$Fe$_{20}$ ferromagnetic layers (15 nm thickness) were grown by e-beam evaporation through a shadow mask on top of OTP thin films. Without breaking the vacuum, the fabricated structures were transferred with another shadow mask back to the first deposition chamber for e-beam evaporation of a SiO$_2$ (300 nm) dielectric layer, and a top Cu thin film (30 nm). The active device area was 0.7 mm×1.0 mm.

### (ii) OTP-SV device fabrication

The half-metal FM LSMO (100 nm) thin films were grown on SrTiO$_3$ (001) substrates by the d.c. magnetron sputtering technique[36], and fabricated for bottom electrodes using wet-etch optical lithography. The LSMO films were cleaned and re-used multiple times since their substrates are without serious degradation. The OTP layer was spin coated on the LSMO electrodes. The fabricated structures were transferred back to the metal deposition chamber for e-beam evaporation of the top FM electrode (Cobalt, 10 nm) and the Au top electrode (50 nm) in a crossbar configuration using a shadow mask. The typical device area was ~ 200 × 200 μm.

### (iii) p-ISHE measurements.

The p-ISHE measurements were made in a Bruker ElexSys E580 X-band (~9.7 GHz) pulsed EPR spectrometer equipped with a dielectric resonator (Bruker FlexLine ER 4118 X-MD5)[23]. The MW pulse duration time was set to 2 μs at a repetition rate of 500 Hz. The maximum pulsed MW power was ~1 kW resulting in an excitation field amplitude $B_1$=1.1 mT at the sample location. The p-ISHE responses were detected by a Stanford Research SRS 570 current-preamplifier, with bandwidth setting of 100Hz-1MHz. The sensitivity of the current-preamplifier was set at 20 µA/V. The p-ISHE($B$) response measurements required averaging over 10240 pulse shots. The parallel capacitance and resistance in the devices were measured using an Agilent E4980A LCR meter.

**(iv) MR measurements.**

The transport measurements were performed in a closed-cycle refrigerator with T in the range from 10 to 300 K, using a Keithley 2400 source meter. The magnetic field, $B$, was applied parallel to the device substrate. The MR is defined as: MR($B$) = ($R(B)$ - $R(0)$)/$R(0)$, where $R(0)$ is the junction resistance at $B = 0$, and $R(B)$ is the resistance measured at field $B$ using the four-points method.

**Acknowledgements**

This work was supported by the DOE, Office of Science, grant DE-SC0014579 (device fabrication, evaluation, and data processing). We also acknowledge the NSF (DMR-1404634) for supporting the development of the ISHE device structures, and the NSF-MRSEC (DMR-1121252) program at the University of Utah for supporting the device fabrication facility.

**Author contributions**

D.S., Z.V.V., and C.B. conceived this study and the experiments. D.S. and C.Z. fabricated the devices. M.K., K.J.S., and H.M. measured the p-ISHE. C.Z. and D.S. measured OTP-SV. C.Z.,

D.S. and M.K. measured the device conductivity and capacitance. M.G. measured the FMR. R.M. performed the Sagnac measurement. C.B. and Z.V.V. were responsible for the project planning, group managing, and manuscript final writing. All authors discussed the results, worked on data analysis and manuscript preparation.

**Additional information**

Supplementary Information is available in the online version of the paper. The authors declare no competing financial interests. Correspondence and requests for materials should be addressed to C. B. (boehme@physics.utah.edu) or Z.V.V. (val@physics.utah.edu).

**Figures and Legends:**

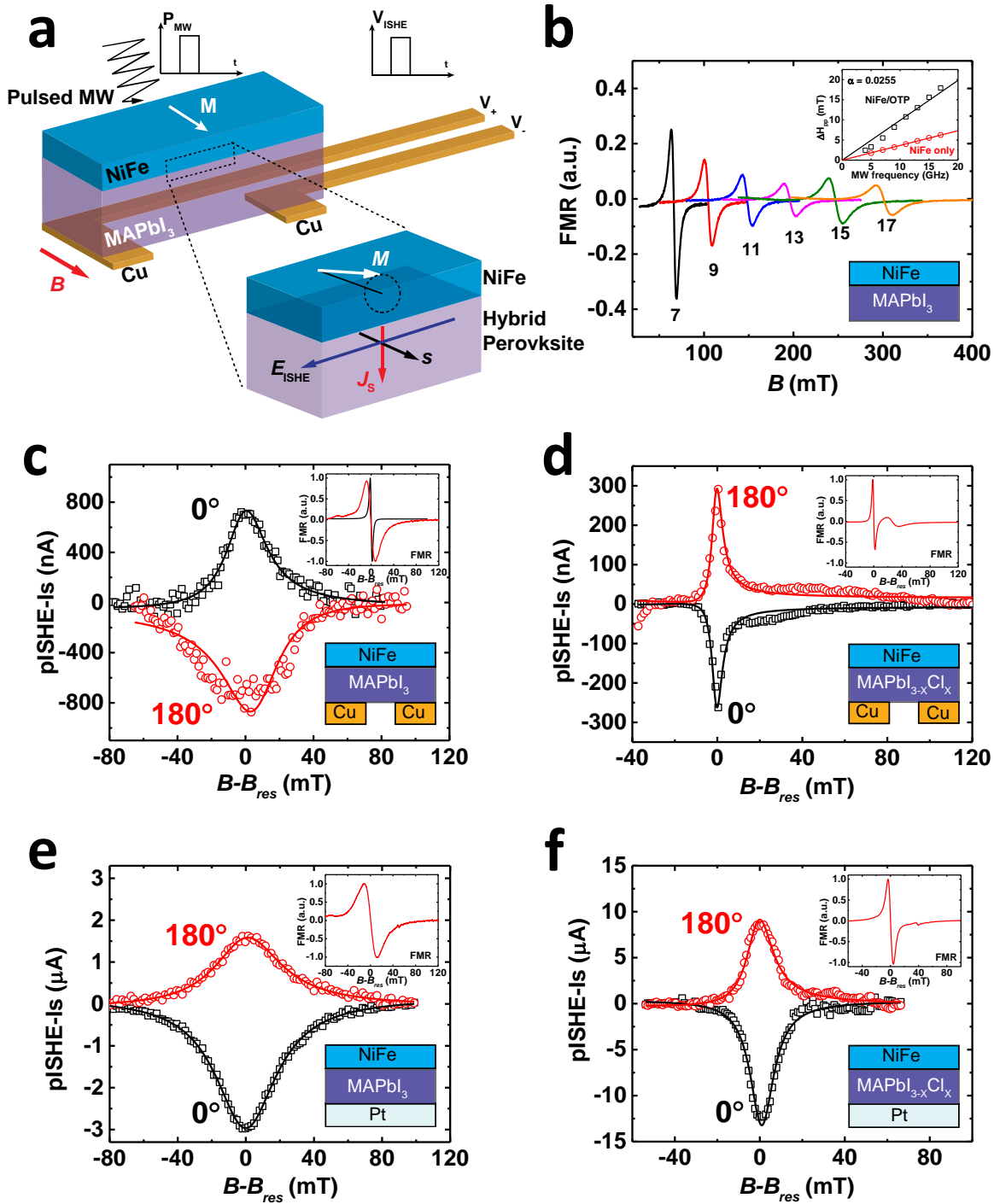

**Figure 1 | Detection of the pulsed ISHE response in OTP thin film. a**, Schematic illustration of the NiFe/OTP/Cu device used for ISHE measurements (upper panel). ***B*** and ***M*** denote, respectively the static external magnetic field and the dynamic magnetization in the NiFe film that precesses around ***B*$_0$*** (lower panel). ***J*$_S$**, ***S***, ***E*$_{ISHE}$**, and ***I*$_{pISHE}$** denote, respectively, the flow of the pulsed spin

current in the OTP layer, its spin polarization vector, generated p-ISHE voltage, and detected p-ISHE current. **b**, Frequency dependence of the FMR($B$) response for the NiFe/OTP bilayer. The inset shows the frequency dependence of the FMR peak-to-peak linewidth and the extracted Gilbert damping factor ($\alpha$) of the NiFe/OTP (10nm) bilayer and NiFe thin film. The black/red solid lines are the linear fit to the data. **c** to **f**, Typical field response of $I_{pISHE}$ (pISHE-$I_S$) measured in various OTP spintronics devices; The black squares and red circles stand for the obtained response using an in-plane magnetic field ***B*** at 0º and 180º, respectively. (**c**) NiFe/MAPbI$_3$/Cu; (**d**) NiFe/MAPbI$_{3-x}$Cl$_x$/Cu; (**e**) NiFe/MAPbI$_3$/Pt; and (**f**) NiFe/MAPbI$_{3-x}$Cl$_x$/Pt. The lines through the data points are fits using a weighted combination of a symmetric ISHE($B$) function and an asymmetric AHE($B$) function (see text). The lower insets are schematics of the corresponding devices. The upper inset in **c** shows the FMR($B$) response of the NiFe film measured without (black) and with (red) the OTP overlayer; whereas the upper insets in **d**-**f** show the FMR($B$) response in the appropriate device.

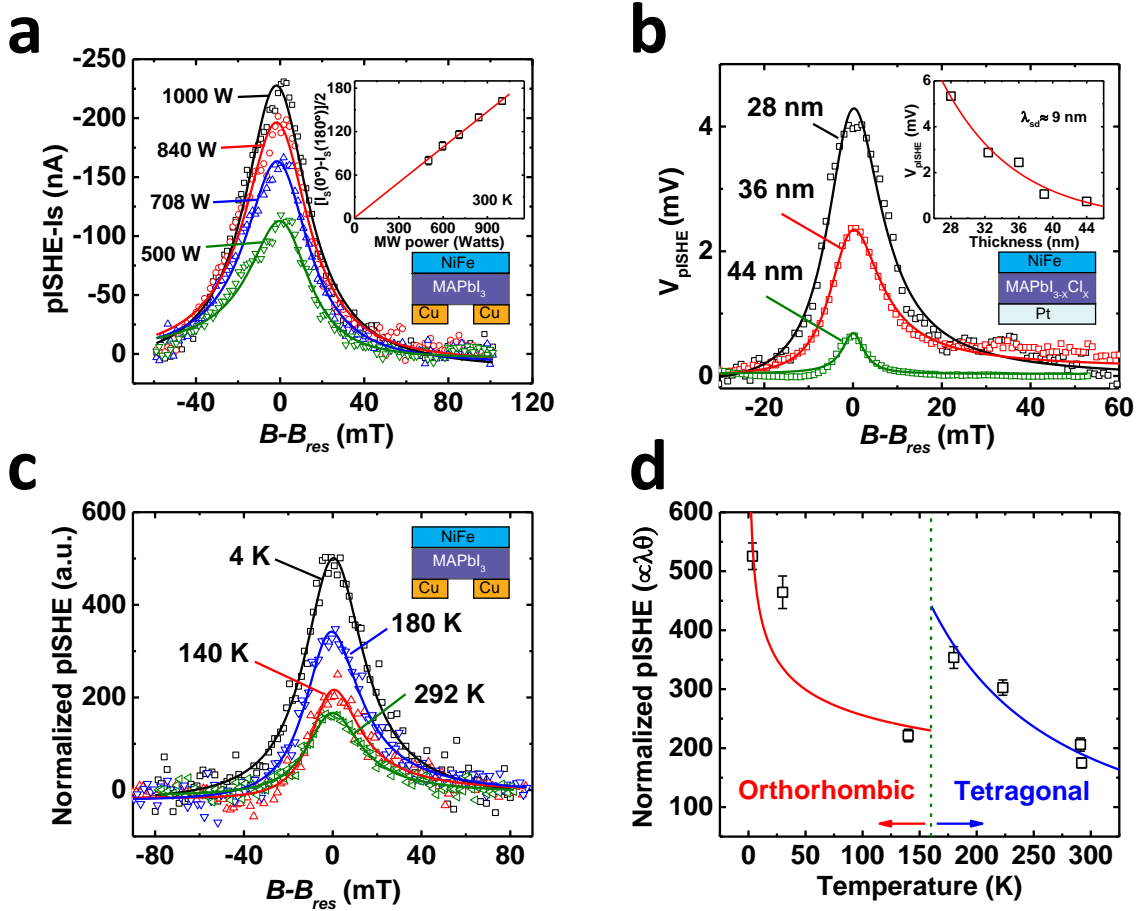

**Figure 2 | MW power, OTP thickness, and temperature dependence of the pISHE(*B*) response in various OTP devices. a**, The room temperature pISHE(*B*) response of NiFe/MAPbI$_3$(25nm)/Cu device as a function of the applied pulsed MW peak power, $P_{MW}$ at *B* around the resonant field, $B_{res}$. The inset summarizes the p-ISHE response vs. $P_{MW}$ with a linear fit through the data points (see S.I. Fig. S3). **b**, The room temperature $V_{pISHE}(B)$ responses of various NiFe/MAPbI$_{3-x}$Cl$_x$/Pt devices at $P_{MW}$ = 1 kW having various OTP thicknesses, $d_N$, as given. The inset summarizes the integrated $V_{pISHE}(B)$ vs. *d*, fitted with an exponential function from which we extracted the spin diffusion length, $\lambda_{sd}$=9nm. **c**, The p-ISHE(*B*) response of NiFe/MAPbI$_3$(10nm)/Cu at various temperatures measured at $P_{MW}$ = 1kW. **d**, The normalized pISHE amplitude (~$\lambda_{sd}\theta_{SH}$) vs. *T* that summarizes the data in **c**. The vertical line is drawn at the tetragonal to orthorhombic phase transition temperature in the MAPbI$_3$ film, which separates the data points in blue (tetragonal) and red (orthorhombic). The lines through the data points are to guide the eye.

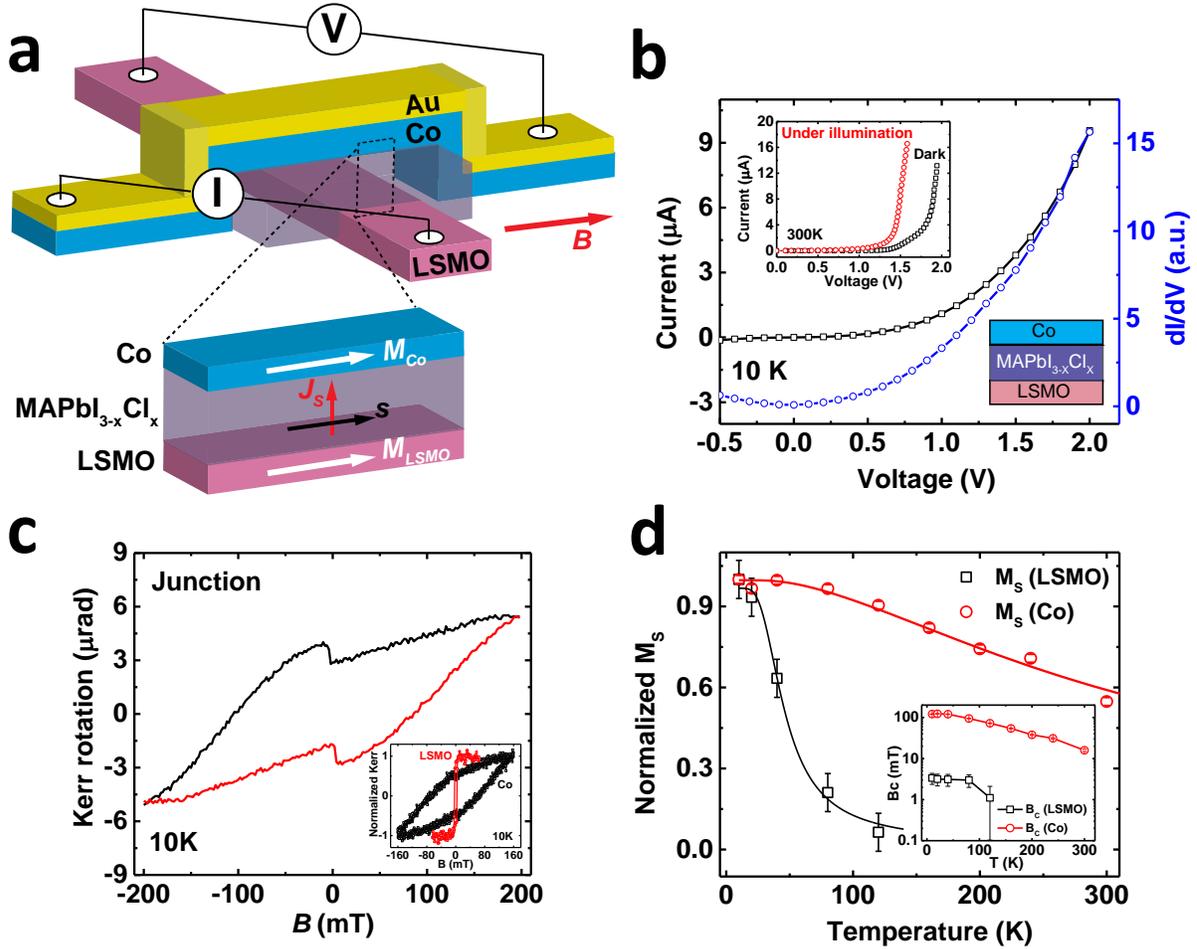

**Figure 3 | Schematic structure of OTP-based spin-valve device and device characterization.**
**a**, Schematic of a perovskite spin-valve (OTP-SV) device based on a LSMO/MAPbI$_{3-x}$Cl$_x$/Co/Au structure. *B* and *M* are the magnetic field and the magnetization of the FM electrodes, respectively. *J$_S$* and *S* are the injected spin-current from the FM electrode and spin polarization vector, respectively. **b**, *I-V* and *dI/dV-V* characteristic response of an OTP-SV device measured at 10K that shows no zero-bias anomaly. The top inset shows the *I-V* characteristic at room temperature measured in the dark (black circles) and upon illumination (red circles). The bottom inset shows the OTP-SV layer structure. **c**, Normalized Kerr rotation angle hysteresis-type response, *θ*(*B*), for the FM electrodes used in the OTP-SV device, measured using MOKE. The inset shows the hysteresis-type *θ*(*B*) measured through the OTP-SV device junction. **d**, The temperature dependence of the saturated Kerr rotation angle *θ* of the two FM electrodes measured through the device junction. The lines through the data points are to guide the eye. The inset shows the temperature dependence of the coercive field for the two FM electrodes measured by MOKE.
18

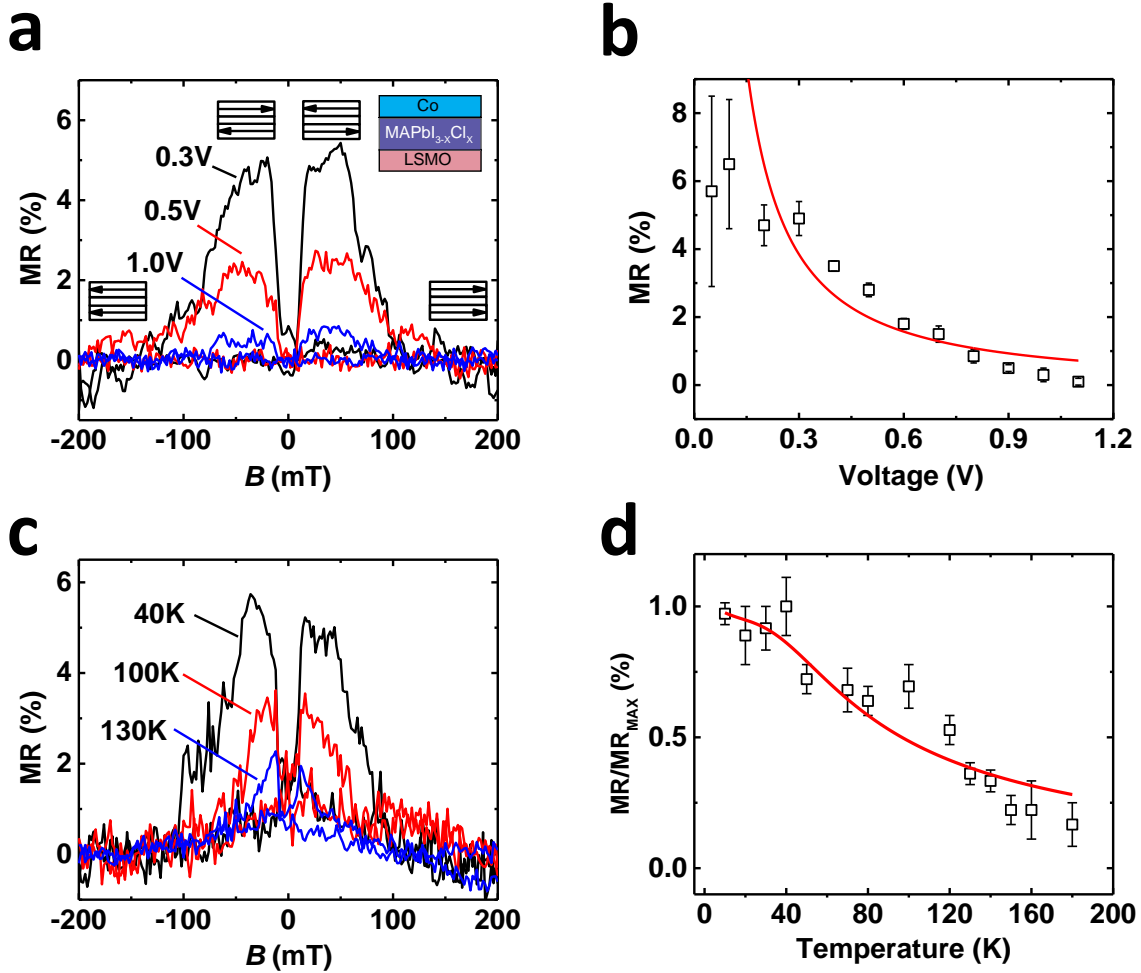

**Figure 4. | Voltage, OTP thickness, and temperature dependence of the MR(*B*) response in OTP-SV devices. a,** MR(*B*) response of LSMO/MAPbI$_3$(245nm)/Co/Al OTP-SV device at various bias voltage, *V*, measured at *T*=10K. The respective parallel (P) and anti-parallel (AP) FM electrode magnetization configurations are also shown. **b,** MR$_{max}$ vs. *V* extracted from the MR(*B*) response. The solid line through the data points is a fit using a power-law, $V^{-p}$, with p = 1.1. **c,** MR(*B*) response of device A at three different temperatures while maintaining a constant current of 100 nA. **d,** Normalized MR$_{max}$ vs. *T*. The solid line through the data points is calculated using Eq. (3), $M_{Co}(T)$ and $M_{LSMO}(T)$ responses from Fig. 3d, and $\lambda_{sd}(T)$ from Fig. 2d.

# Spintronics of Organometal Trihalide Perovskites

Dali Sun†, Chuang Zhang†, Marzieh Kavand†, Kipp J. van Schooten, Hans Malissa, Matthew Groesbeck, Ryan McLaughlin, Christoph Boehme*, and Z. Valy Vardeny*

*Department of Physics & Astronomy, University of Utah, Salt Lake City, Utah, 84112, USA*

†These authors contributed equally to this work.
*To whom correspondence should be addressed: boehme@physics.utah.edu, val@physics.utah.edu

**Supplementary Materials:**

**A. Materials and Methods**

    **(i)**    **Solution-based fabrication of OTP films**

The perovskite precursor solution was made by mixing 795 mg of $CH_3NH_3I$ (Dyenamo) with 231 mg of $PbI_2$ (Sigma Aldrich) and 417 mg of $PbCl_2$ (Sigma Aldrich) in 2.5 mL DMF (N,N-dimethylformamide) solvent. The Cl-contained precursor was adopted to slow down the perovskite crystallization thereby forming a uniform and smooth polycrystalline film. The solution was diluted down to a concentration of 0.2 M to prepare thinner perovskite layers. The obtained solution was left for stirring overnight on a hotplate at 70°C in the glove box under nitrogen atmosphere ($O_2$/$H_2O$<0.5 ppm). Once cooled to room temperature, the bright yellow solution was spin coated on the substrate (at 2000 to 8000 RPM). These films turned their color to dark brown after annealing at 90°C for 15 minutes and then 105°C for 120 minutes, indicating the formation of $CH_3NH_3PbI_{3-x}Cl_x$ perovskite layer (Fig. S1-S2). The thickness of OTP layers was measured by a profilometer (KLA Tencor).

    **(ii)**    **Co-deposited preparation of OTP films**

To thermally evaporate $CH_3NH_3PbI_3$ films[5], two crucibles with $PbI_2$ (Sigma Aldrich) and $CH_3NH_3I$ (Dyenamo) powders were placed separately in the deposition chamber, devoted for organic deposition (base pressure $1\times10^{-7}$ Torr). In a typical preparation, $PbI_2$ and $CH_3NH_3I$ were first degassed and then heated up to ~290 °C and ~90 °C respectively for thermal evaporation. The growth rate was controlled at 0.2 Å/s for $PbI_2$ and 1.0 Å/s for $CH_3NH_3I$ with a volume ratio of 1:5, monitored by a quartz crystal thickness monitor. The sample holder was continuously rotating at

~3 RPM and heated to 70 °C (in situ annealing) during the deposition process. Following the deposition procedure, the films were post-annealed at 70 °C in the chamber for 30 mins (see Fig. S1-S2).

### (iii) Measurement procedures

**X-ray Diffraction (XRD), Absorption (Abs) and Photoluminescence (PL) measurements.** The XRD of $MAPbI_{3-x}Cl_x$ and $MAPbI_3$ layers on glass substrates were measured by PANalytical X'Pert X-Ray Diffractometer. The absorption and the steady-state PL emission were measured by using LAMBDA 950 UV/VIS/NIR and a spectrometer (Ocean-optics, USB4000-UV-VIS), respectively.

**Photovoltaic power conversion efficiency (PCE) measurements.** The current-voltage and PCE characteristics of the OTP-PV cells were measured using a computer-controlled source-measurement unit (Keithley 236) and Xenon lamp with a AM 1.5G filter inside the glove box, calibrated by a standard silicon PV cell.

**FMR measurements.** The FMR($B$) response as a function of MW frequency (4 GHz to 17 GHz) was performed using a home-made coplanar waveguide and a MW source generator (Keysight, N5173B EXG MW Generator). The MW frequency dependence of the resonant field ($B_{res}$) was fitted by the Kittel formula ($f = \frac{\gamma}{2\pi}\sqrt{B_{res}(B_{res} + 4\pi M_S)}$) to derive the $g$-factor and effective magnetization ($4\pi M_S$). The damping factor for the OTP/NiFe junction is fitted from the FMR peak-to-peak width ($\Delta H_{pp}$) as a function of the MW frequency[30]: $\Delta H_{pp} = \Delta H_0 + \frac{4\pi\alpha}{\sqrt{3}\gamma}f$, where $\gamma$ is the gyromagnetic ratio extracted from the obtained $g$ factor. $\Delta H_0$ denotes the inhomogeneous broadening.

**Sagnac MOKE measurements.** Magneto-optics Kerr effect (MOKE) sensitive Sagnac interferometer was constructed based on the design conceived by J. Xia et. al[40]. The technical details of the interferometer are described in ref. 40. Our home-built Sagnac interferometer has a static DC Kerr rotation sensitivity of ~20 nRad. The Sagnac interferometer was installed onto a 2D scanning stage with a spot size of ~100 micron and a maximum image size of 25mm × 25mm in order to spatially measure the magnetizations of the individual FM electrode or junction in the OTP-SV device.

## B. OTP thin film and PV device characterizations

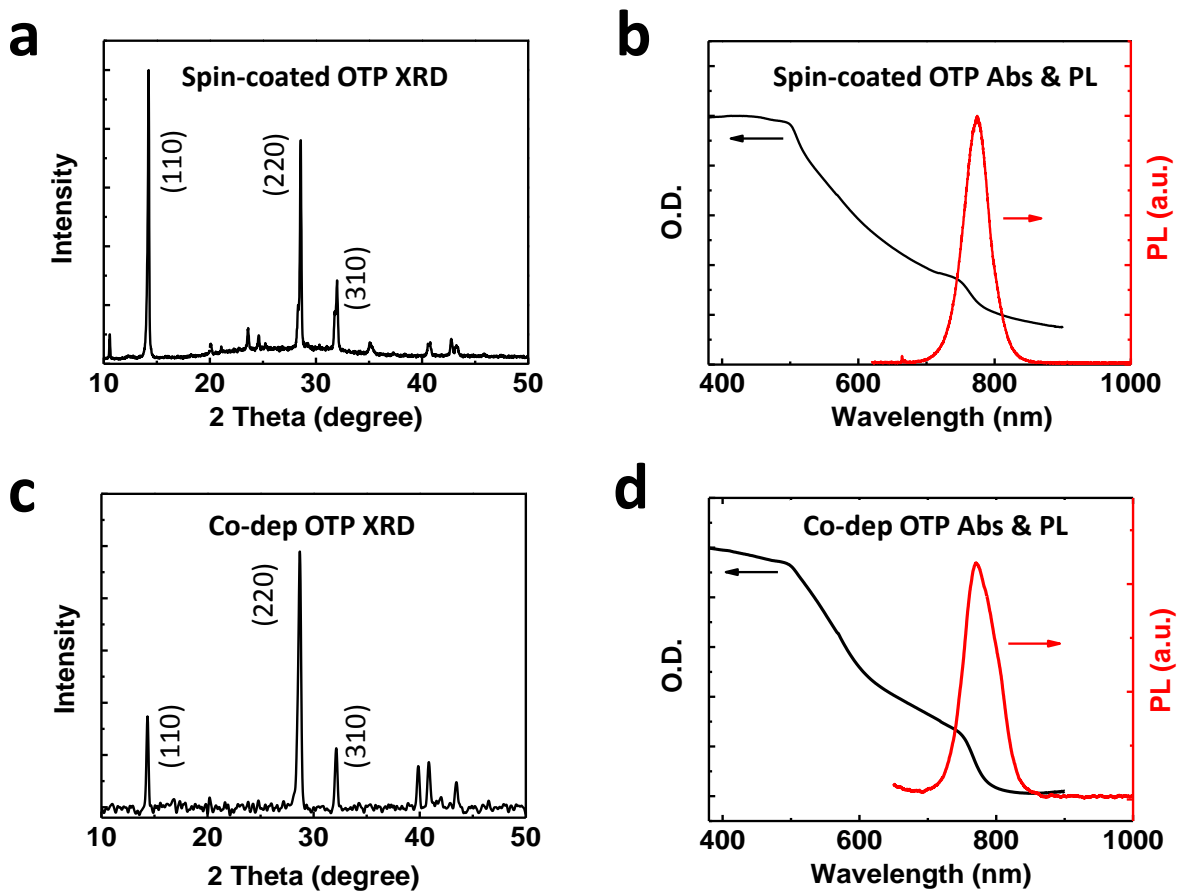

**Fig. S1.** Spin-coated and co-deposited OTP layer characterization. **a** and **c**, XRD spectra of a MAPbI$_{3-x}$Cl$_x$ film (spin coated) and MAPbI$_3$ (co-deposited) film, respectively. The main diffraction peaks (i.e. 110, 220, and 310) are identical for both films. **b** and **d**, the absorption and steady-state PL emission spectra of MAPbI$_{3-x}$Cl$_x$ and MAPbI$_3$ film, respectively. Both films show similar XRD, absorptions and PL spectra, in good agreement with literature reports[1-13].

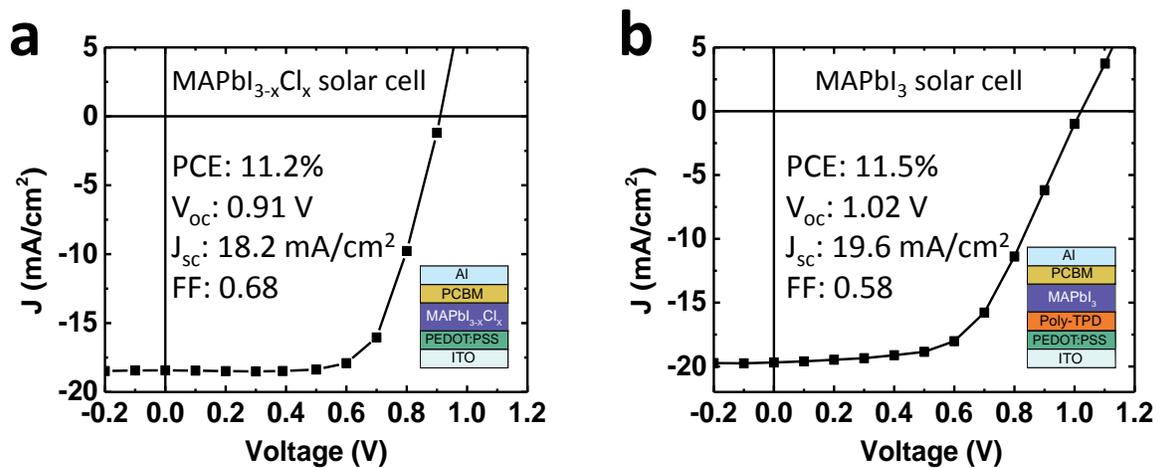

**Fig. S2.** Typical *J-V* characteristics of MAPbI$_{3-x}$Cl$_x$ (spin coated, **a**) and MAPbI$_3$ (co-deposited, **b**) based solar cells under standard AM 1.5G illumination. The obtained PCEs are closed to the values of the optimized OTP based solar cells using the same structure (~12%, Ref. 9). Insets show schematic illustration of the solar cell device structure.

## C. Heating effects during p-ISHE measurement in OTP spin-pumping devices

The obtained p-ISHE($B$) response is the combination of the actual ISHE in the OTP film, the anomalous Hall effect (AHE) in the NiFe film, and a heating effect. The contribution of AHE voltage can be separated by fitting the p-ISHE($B$) response using a combination of symmetric (absorption shape, ISHE) and asymmetric (dispersion shape, AHE) functions ($I_{pISHE}(B) = I_S(B) + I_{AHE}(B)$). To separate the contribution of heating to the p-ISHE response, we define $act\_ISHE = (I_S(0°) - I_S(180°))/2$, because the actual ISHE response reverses polarity when the $B_0$ direction is reversed, whereas the signal arising from the heating effect (heat($B$)) is independent on the $B_0$ direction[25], given by: $heat(B) = (I_S(0°) + I_S(180°))/2$. The MW power dependence of the obtained ISHE($B$) response for the NiFe/CH$_3$NH$_3$PbI$_3$/Cu device is shown in Fig. 2a and Fig. S3. Panels (c) and (d) show the MW power dependence of actual ISHE and heating signals, respectively. After separating the heating effect, the actual ISHE response increases linearly with $P_{MW}$ as expected for the ISHE response induced by the spin pumping, whereas the heating signal shows a non-linear increase as a function of $P_{MW}$.

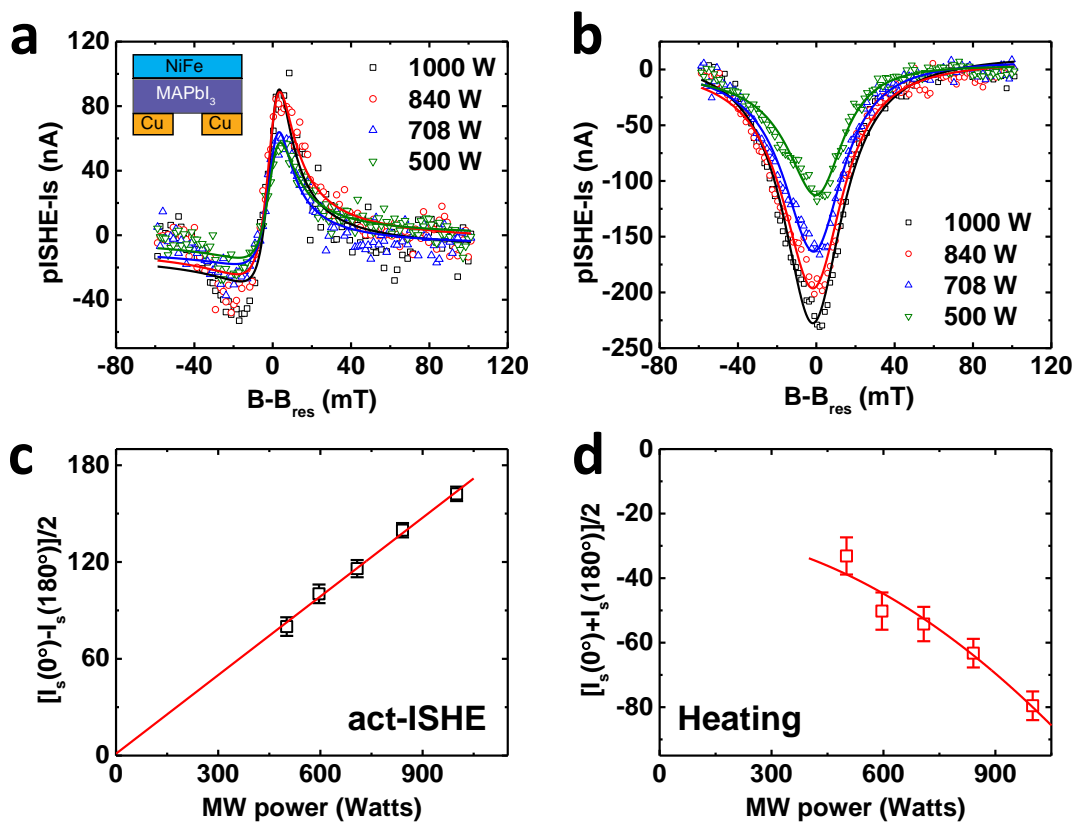

**Fig. S3. a** and **b,** MW power dependence of the p-ISHE($B$) response in the MAPbI$_3$ based ISHE device with the in-plane magnetic field $B$ (at 0º) and $-B$ (at 180º), respectively. The inset shows schematically the MAPbI$_3$ ISHE device geometry. **c** and **d**, the obtained actual-ISHE and heating signals as a function of MW power, respectively. The red lines through the data present the linear **c** and power law fitting **d**.

## D. The p-ISHE measurements of Cu/MAPbI$_{3-x}$Cl$_x$/NiFe device

To elucidate the influence of chlorine incorporation on the polarity of p-ISHE response, a MAPbI$_{3-x}$Cl$_x$ device with an 'inverted structure' was investigated. A NiFe layer was first deposited onto the FM substrate, followed by the spin-coated MAPbI$_{3-x}$Cl$_x$ layer and two Cu contacts, as shown in the bottom inset of Fig. S4. Recent studies show that the chlorine additive atoms only stay close to the MAPbI$_{3-x}$Cl$_x$ film surface[31]. In this 'inverted structure' the NiFe FM is located at the bottom of the MAPbI$_{3-x}$Cl$_x$ film, whereas no chlorine atoms were found inside the film, as confirmed by XPS analysis (Ref. 31). Under these conditions, the spin-pumping device is expected to have the same polarity of p-ISHE response as that of MAPbI$_3$ based device, although the polarity is reversed because the spin-current direction is reversed in the 'inverted structure' ($ISHE \propto \theta_{SHE} \boldsymbol{J}_S \times \boldsymbol{S}$). Indeed, the observed p-ISHE response in the 'inverted structure' of MAPbI$_{3-x}$Cl$_x$ device shows a negative sign at 0 degree, having the same polarity as that in the normal structure when NiFe is deposited on top of MAPbI$_3$ film. We thus conclude that these observations are consistent with the previous reports that chlorine sits only on the surface of MAPbI$_{3-x}$Cl$_x$ film[31].

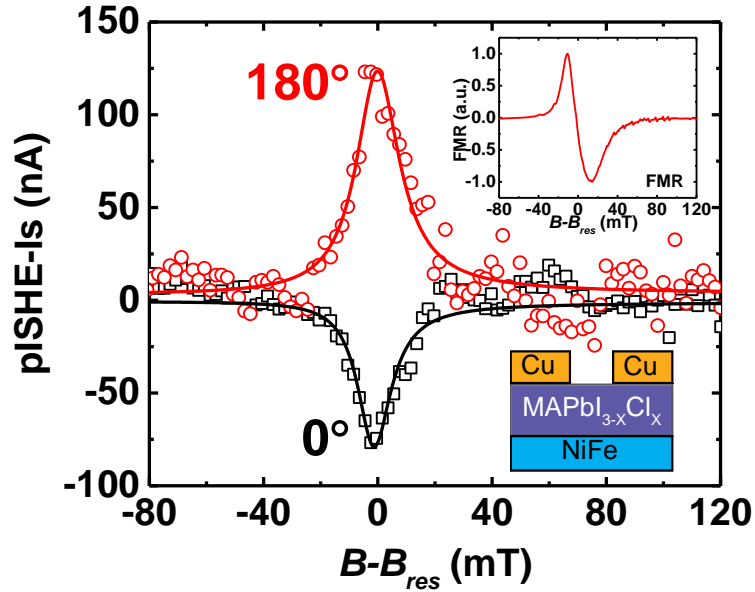

**Fig. S4.** p-ISHE(B) response for the MAPbI$_{3-x}$Cl$_x$ (130 nm) based device with an 'inverted structure'. The p-ISHE($B$) responses are fitted by the combination of ISHE(B) and AHE(B) response components (solid lines through the open symbols). The black squares and red circles lines are the data with the in-plane magnetic field $\boldsymbol{B}$ (at 0º) and $-\boldsymbol{B}$ (at 180º), respectively. Bottom inset is the schematic of the 'inverted device' geometry. Top inset shows the FMR($B$) response.

## E. p-ISHE in NiFe/MAPbI3/Pt devices as a function of the OTP thickness

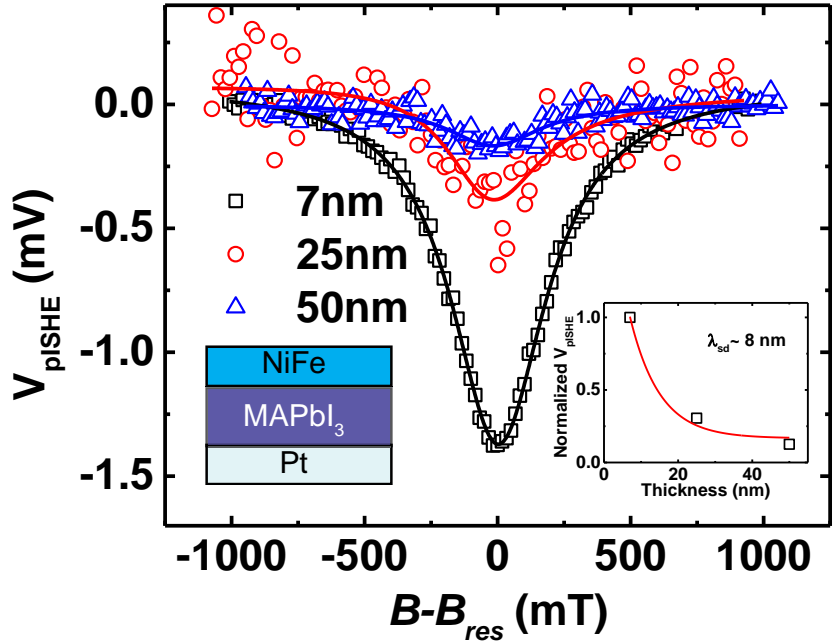

**Fig. S5.** $V_{pISHE}(B)$ response of a trilayer NiFe/MAPbI$_3$/Pt device at $P_{MW}$ = 1 kW as a function of the OTP interlayer thickness. The inset summarizes the obtained $V_{pISHE}$ vs. the OTP film thickness $d$, fitted by an exponential decay, from which the spin diffusion length $\lambda_{sd} = 8\ nm$ is extracted[27].

## F. Estimations of Spin Hall angle in the OTP films from p-ISHE measurements

The spin hall angles ($\theta_{SH}$) of the OTP films are quantified based on a phenomenological model combined with an equivalent circuit model of our set-up[23]. The p-ISHE response is measured during each MW pulse excitation (2 μs duration) at 500 Hz repetition rate. Consequently, the generated p-ISHE response contains a wide bandwidth of AC signals (from ~100 Hz to ~1 MHz, Ref. 23). The weighted capacitance of the OTP film, $C_N$, also needs be considered in the circuit model. By taking into account the device structure, the detection electronics, and the AC response of each electronic component, a simplified expression may be written:

$$I_S(pISHE) = Re[(I_C + I_F)\frac{R_F}{R_S^{SUM}+R_F+\frac{2R_N^{SUM}}{1+i(\omega_j C_{N(j)}R_{N(j)})^{SUM}}}], \tag{S1}$$

where $R_F$ and $R_S^{SUM}$ are the series resistances of the NiFe film and current-preamplifier impedance, respectively. $\omega_j$, $C_{N(j)} = \frac{\varepsilon_{N(j)}w(\frac{l}{2})}{d_N}$ and $R_{N(j)} = \frac{d_N}{\sigma_{N(j)}W(\frac{l}{2})}$, are respectively the j- frequency component (established by the finite Fourier transform), parallel capacitance, and resistance of the OTP layer at $\omega_j$ (see Ref. *21* for the derivation). The variables $\varepsilon_{N(j)}$ and $\sigma_{N(j)}$ are the dielectric constant and conductivity of the OSEC material at $\omega_j$. $R_N^{SUM}$ and $(\omega_j C_{N(j)}R_{N(j)})^{SUM}$ are the respective sum of parallel resistance, and product $\omega C_N R_N$ terms averaged over the entire frequency range of the measurement apparatus. The parameter $w$ is the width of NiFe layer, whereas $l$ is the length of NiFe thin film. $I_C$ and $I_F$ are the generated ISHE responses at the NiFe/OTP interface, and AHE response from the NiFe thin film, respectively. The latter response can be separated by fitting the p-ISHE (B) response by the combination of a symmetric (pure ISHE response) and an asymmetric (dispersion shape, AHE(*B*))) function. The spin-pumping related $I_C$ through the OTP layer can be expressed as:

$$I_C = l\theta_{SH}\left(\frac{2e}{\hbar}\right)\lambda_N \tanh\left(\frac{d_N}{2\lambda_N}\right)j_S^0, \tag{S2}$$

where $\theta_{SH}$ and $\lambda_N$ are the respective spin Hall angle and spin diffusion length in the OTP, and $j_S^0$ is the spin current in the OSEC perpendicular to the NiFe/OTP interface and along $l$[28].

For the analysis of the p-ISHE we need to estimate $j_S^0$. In the model for spin-pumping, the injected spin current density $j_S^0$ at the interface is expressed by the relation[28]:

$$J_S^0 = \frac{g_r^{\uparrow\downarrow}\gamma^2 h^2\hbar\left[4\pi M_S\gamma\sin^2\theta_m + \sqrt{(4\pi M_S)^2\gamma^2 + 4\omega^2}\right]}{8\pi\alpha^2[(4\pi M_S)^2\gamma^2\sin^4\theta_m + 4\omega^2]} \tag{S3}$$

where $\theta_m$ is the magnetization angle to the normal axis of the film plane, $\omega$ is the angular frequency of the magnetization precession (at the MW frequency), $g_r^{\uparrow\downarrow}$ is the mixing conductance, $\gamma$ is the gyromagnetic ratio, $\alpha$ is the Gilbert damping constant, and $M_S$ is the saturation magnetization. $h$ is the $B_1$ field component of the MW excitation ($h^2$ may also be considered to be proportional to the amplitude of FMR absorption). The real part of the mixing conductance is given by[28]:

$$g_r^{\uparrow\downarrow} = \frac{2\sqrt{3}\pi M_S \gamma d_F}{g\mu_B \omega}(\Delta H_{pp_{(NiFe/OTP)}} - \Delta H_{pp_{(NiFe)}}); \tag{S4}$$

Where $g$ is the electron g-factor and $\mu_B$ is the Bohr magneton. $\Delta H_{pp_{(NiFe/OTP)}}$ and $\Delta H_{pp_{(NiFe)}}$ are the FMR spectral peak-to-peak width for NiFe/OTP) and pure NiFe film, respectively. With this, the Lamda-theta product ($\lambda\theta_{SH}$) and thus, the spin Hall angle $\theta_{SH}$ could be calculated (if $\lambda$ is known) by substituting the above parameters into Eqs. (S1) and (S2)[23, 28]. Here we used the values of conductivity ($\sim 1.0 \times 10^{-5} s/m$, Ref. 41-43) and a dielectric constant ($\sim 27$, Ref. 44-46) from literature reports for the estimation of spin hall angles in OTP layers.

## G. Temperature dependence of the p-ISHE response

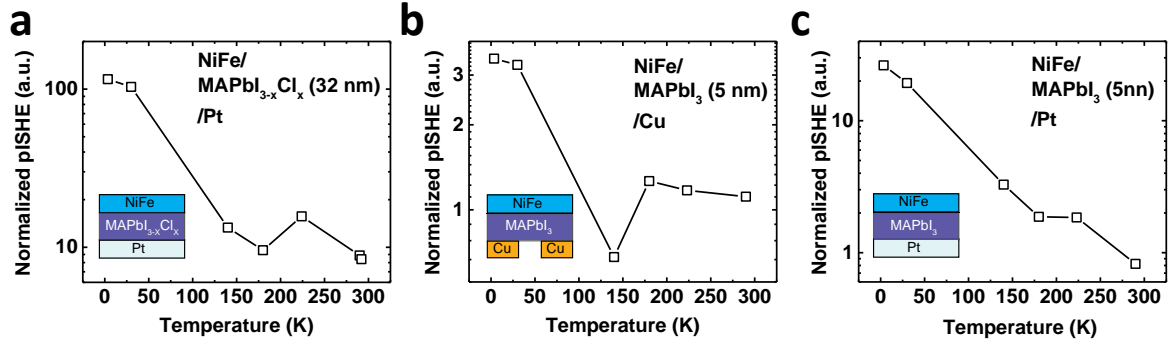

**Fig. S6.** p-ISHE responses of three OTP-ISHE devices as a function of temperature (proportional to the product "$\lambda_{sd}\theta_{SHE}$"), normalized by taking into account the temperature dependence of the FMR width, amplitude and device resistance. $\lambda_{sd}\theta_{SHE}$ for all three devices shows a discontinuity around $T\sim 150$K.

## H. *I-V* characteristics of an OTP-SV device at various temperatures

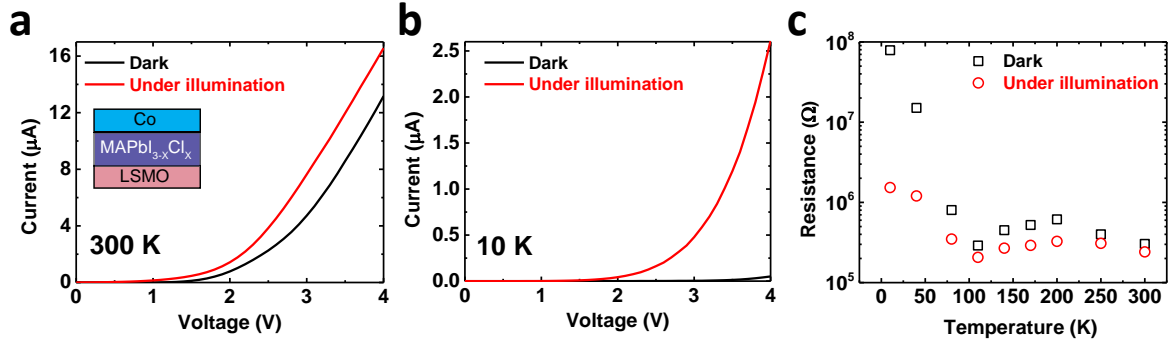

**Fig. S7. a** and **b,** *I-V* characteristics of OTP-SV device with and without illumination at 300K and 10K, respectively. **c**, Device resistance as a function of temperature (at constant bias voltage of 4V) with and without illumination. The resistance is reduced roughly by two orders of magnitude due the photogenerated current upon the illumination at lower temperature, indicating the occurrence of drift-current rather than tunneling in the OTP-SV device.

## I. Ref. 47: Abstract submitted to the 2016 APS March Meeting

### Photoexcited carrier spin dynamics in $CH_3NH_3PbI_3$*

P, Odenthal, N. Gundlach, W. Talmadge, R. Wang, C. Zhang, D. Sun, Z. Valy Vardeny, Y. Li.

Department of Physics & Astronomy, University of Utah, Salt Lake City UT 84112

Metal halide perovskites have shown great promise for the field of spintronics due to their large tunable spin-orbit coupling, spin dependent optical selection rules and predicted electrically tunable Rashba band. The spin sensitive optical transitions allow optical spin orientation of carriers using circularly polarized light, and detection of the spin polarization via optical Faraday rotation measurement. We studied carrier spin dynamics in solution-processed polycrystalline $CH_3NH_3PbI_3$ films using time-resolved Faraday rotation (TRFR). TRFR reveals unexpected long spin lifetimes exceeding 1ns at 4K. This is significant given that Pb and I exhibit large spin-orbit coupling, which usually lead to fast spin relaxation.

* Supported by the NSF-MRSEC program at the University of Utah